\pdfoutput=1
\documentclass[11pt,a4paper]{article}
\usepackage[T1]{fontenc}
\usepackage[utf8]{inputenc}
\usepackage{amsmath}
\usepackage{txfonts}
\usepackage[a4paper,margin=25mm]{geometry}
\usepackage{graphicx,booktabs}
\usepackage[dvipsnames]{xcolor}
\usepackage[font=small,labelfont=bf,labelsep=colon]{caption}
\usepackage{enumitem}
\usepackage{setspace}
\usepackage[american,cuteinductors,smartlabels]{circuitikz}
\usetikzlibrary{positioning,arrows.meta}
\usepackage[colorlinks,linkcolor=blue!70!black,citecolor=blue!70!black,urlcolor=blue!70!black]{hyperref}
\setlist[itemize]{leftmargin=1.2em,itemsep=1pt,topsep=4pt}
\setlist[enumerate]{leftmargin=1.5em,itemsep=1pt,topsep=4pt}
\newcommand{\para}[1]{\par\medskip\noindent\textbf{#1}\quad}
\newcommand{\Ho}{\mathbf{H}_1}
\newcommand{\rc}{\rho_0 cS}
\renewcommand{\Re}{\mathop{\mathrm{Re}}\nolimits}

\newcommand{\pzFour}{0.139}
\newcommand{\pzSeven}{0.038}
\newcommand{\pzTen}{0.021}
\newcommand{\pzThirteen}{0.014}
\newcommand{\pzSixteen}{0.010}
\newcommand{\pzThirteenFive}{4.3\times10^{-4}}

\newcommand{\pzTwentytwoFive}{1.2\times10^{-5}}

\newcommand{\ladFourteenFive}{4.3\times10^{-4}}
\newcommand{\tauOpt}{0.655}

\newcommand{\hornRt}{25.4}
\newcommand{\hornRm}{30}
\newcommand{\hornL}{0.8}
\newcommand{\hornM}{6.17}
\newcommand{\hornFc}{168}
\newcommand{\hornAreaRatio}{140}
\newcommand{\hornKaFc}{0.93}
\newcommand{\hornKaMin}{0.27}
\newcommand{\hornKaMax}{27}
\newcommand{\hornMatchedCheck}{2\times10^{-14}}
\newcommand{\cfInfStop}{10^{-15}}
\newcommand{\cfInfPass}{0.92}
\newcommand{\cfInfPassMax}{20}
\newcommand{\cfInfOff}{0.1}
\newcommand{\cfInfOffEighty}{0.007}
\newcommand{\stHundred}{0.012}
\newcommand{\ladHundred}{0.13}

\newcommand{\stThousand}{1.2\times10^{-4}}
\newcommand{\ladThousand}{0.0010}
\newcommand{\hornLambdaRatio}{12}
\newcommand{\stOrder}{2.0}
\newcommand{\ladOrder}{2.0}
\newcommand{\stNeeded}{344}
\newcommand{\ladNeeded}{1020}
\newcommand{\lossMax}{0.013}
\newcommand{\lossFreq}{202}
\newcommand{\lossHigh}{0.003}

\newcommand{\htrhoc}{0.986}

\newcommand{\htber}{0.155}

\newcommand{\htlaFour}{0.187}

\newcommand{\htlbFour}{0.102}

\newcommand{\htlbFourteen}{0.023}
\newcommand{\hornSens}{1.06}
\newcommand{\totalElements}{2055}

\begin{document}

\begin{center}
{\LARGE\bfseries Continued fractions and passive ladder networks for acoustic impedances:
the baffled piston and the horn\par}
\vspace{5mm}
{\large Ronald M. Aarts\par}
\vspace{2mm}
{\small Eindhoven University of Technology, Department of Electrical Engineering,\\
P.O. Box 513, 5600 MB Eindhoven, The Netherlands\\
r.m.aarts@tue.nl\par\smallskip
\textit{Preprint; to be submitted to Acta Acustica}\par}
\end{center}
\vspace{4mm}

\noindent\textbf{Abstract.}\quad
Lumped-element networks give transparent, passive models of acoustic
impedances whose low- and high-frequency limits can be read from the circuit, but a systematic,
arbitrarily accurate construction for non-rational impedances is lacking.
Continued fractions, one- and two-point Pad\'e approximants and their identity
with ladder networks and Cauer synthesis are introduced in tutorial form and applied to two
examples: the radiation impedance of a baffled circular piston and the throat impedance of a horn
with a piston mouth.
The even contraction of the piston's continued fraction has positive
coefficients, certified by interval arithmetic for the first 100 pairs, and yields a ladder of
positive resistances, masses and compliances; with 14 elements the error is $4\times10^{-4}$ up to
$ka=5$. Beranek's four-element network is, to about 2\%, the order-two two-point Pad\'e approximant. A
horn modelled as a cascade of waveguide sections is a continued fraction whose tail is the mouth
impedance; the infinite exponential horn is a periodic continued fraction, an exact high-pass ladder.
The piston networks are passive at every certified order (up to 202 elements)
and are extended without changing existing elements; for a finite horn the error of the mouth network passes one to one to
the throat impedance.


\bigskip
\section{Introduction}\label{sec:intro}

Lumped-element models, built from masses, compliances and resistances, may look old-fashioned
next to finite-element and boundary-element methods, but they are not. A network makes the physics
visible: at low frequency the masses and compliances closest to the terminals dominate, at high
frequency only the resistances remain, and both limits can be read directly from the circuit diagram.
A network of positive elements is passive by construction, combines without effort with the
electrical and mechanical parts of a transducer model, and runs unchanged in circuit simulators and
in the time domain. What is missing is a systematic way to obtain such networks for impedances that
are not rational, such as radiation impedances, with an accuracy that can be raised at will.

This paper provides that link. It introduces, in tutorial form, the mathematics behind it:
continued fractions, Pad\'e and two-point Pad\'e approximants, and their identity with ladder
networks and Cauer synthesis. It then applies them to two acoustic examples: the radiation
impedance of a baffled circular piston, and the throat impedance of a horn terminated by that
piston.

A power series is the most natural local description of an analytic function, but a poor global
one: it converges only in a disk that stops at the nearest singularity, and a truncated series is
a polynomial, which cannot reproduce poles, branch cuts or a finite limit at infinity. A continued
fraction uses the same local information but packages it as a sequence of rational functions.
Rational functions can have poles and level off at infinity, which is why continued fractions are
the workhorse behind many library routines for special functions~\cite{press,cuyt}.

Rational functions are also what passive electrical and mechanical networks produce. A network of
positive resistances, inductances and capacitances (or mechanical resistances, masses and
compliances) has a positive-real driving-point impedance, and conversely every rational
positive-real impedance can be realised with such elements: Foster~\cite{foster} and
Cauer~\cite{cauer} gave canonical realisations of lossless and other two-element-kind impedances,
Cauer's in ladder (continued-fraction) form, Brune~\cite{brune} treated the general case with ideal
transformers, and Bott and Duffin~\cite{bottduffin} removed the transformers. Bott--Duffin
networks, however, contain many more energy-storage elements than the McMillan degree of the
impedance, an overhead shown only recently to be necessary for particular
classes~\cite{hughessmith2014,hughes2017}. Kalman argued that network synthesis deserves renewed
mathematical attention~\cite{kalman2010}, and Smith's introduction of the inerter~\cite{smith2002}
revived the subject for mechanical networks.

The case study of this tutorial is of a different kind: not the realisation of a given rational
function, but the construction of positive-real approximants of a \emph{transcendental} one, the
radiation impedance of a baffled piston~\cite{rayleigh}
\begin{equation}\label{eq:Z}
\frac{Z(\omega)}{\rc}=1-\frac{2J_1(2ka)}{2ka}+\mathrm{j}\,\frac{2\Ho(2ka)}{2ka},
\end{equation}
with $S=\pi a^2$, $J_1$ the Bessel and $\Ho$ the Struve function and time convention
$e^{\mathrm{j}\omega t}$. It enters nearly every lumped model of a loudspeaker or other baffled
transducer, and circuit and state-space simulators need it as a finite network. Beranek's
four-element network~\cite{beranek,leach}, \cite[Fig.~4.37, pp.~200--201]{beranekmellow2019} is the
classical choice; it is correct at low and high frequencies but misses the overshoot of $\Re Z$
near $ka\approx2.5$ and the subsequent ripple. Ideally one wants a network that is
\begin{itemize}
\item built from positive, constant R, M and C only;
\item passive at every order without any check;
\item extendable to any order without changing the elements already present.
\end{itemize}
A family with the last property is called \emph{nested} here.

\para{Main findings.}No single representation is best for every purpose. The nested ladder
constructed here is the choice when nesting, a construction without optimisation, or very high
accuracy up to a prescribed $ka$ matters; adapting only its last one or two elements (a broadband
termination) reduces the broadband error with 14 elements from 0.068 to 0.023. A fitted, non-nested
Foster-type network is somewhat more accurate over a wide band at low order. At four elements a
closed-form two-point Pad\'e network, practically identical to Beranek's, is close to the best
possible. When no network is needed at all, a direct approximation of the Struve function is far
more accurate than any of them.

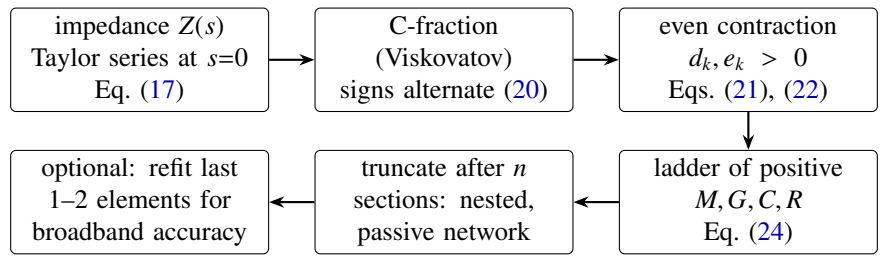
\begin{figure}[htbp]
\centering
\begin{tikzpicture}[node distance=5mm and 6mm,
  box/.style={draw,rounded corners=2pt,align=center,font=\small,minimum height=12mm,text width=32mm,inner sep=3pt},
  arr/.style={-{Stealth[length=2mm]},thick}]
\node[box] (z) {impedance $Z(s)$\\Taylor series at $s{=}0$\\Eq.~\eqref{eq:taylor}};
\node[box,right=of z] (c) {C-fraction\\(Viskovatov)\\signs alternate~\eqref{eq:cfrac}};
\node[box,right=of c] (j) {even contraction\\$d_k,e_k>0$\\Eqs.~\eqref{eq:jfrac}, \eqref{eq:de}};
\node[box,below=of j] (l) {ladder of positive\\$M,G,C,R$\\Eq.~\eqref{eq:elements}};
\node[box,left=of l] (t) {truncate after $n$\\sections: nested,\\passive network};
\node[box,left=of t] (b) {optional: refit last\\1--2 elements for\\broadband accuracy};
\draw[arr] (z) -- (c); \draw[arr] (c) -- (j); \draw[arr] (j) -- (l); \draw[arr] (l) -- (t); \draw[arr] (t) -- (b);
\end{tikzpicture}
\caption{The route followed for the piston: from the power series of the impedance to a nested
ladder of positive elements. Positivity of $d_k,e_k$ is certified for $k\le100$
(Sec.~\ref{sec:pos}); the optional termination is described in Sec.~\ref{sec:term}.}\label{fig:route}
\end{figure}

\para{Plan.}Sections~\ref{sec:nutshell}--\ref{sec:twopoint} introduce continued fractions, their
construction, their convergence and two-point Pad\'e approximants, with worked examples.
Section~\ref{sec:networks} connects them to ladder networks. Section~\ref{sec:piston} applies
everything to the piston: a nested ladder whose construction extends to arbitrary order, with a
computer-assisted positivity proof up to 202 elements, its accuracy
in frequency and time, its broadband termination, a one-parameter family of four-element networks
that contains Beranek's, and a comparison of all methods. Section~\ref{sec:horn} treats the horn,
whose throat impedance turns out to be a continued fraction ending in the piston impedance.
Figure~\ref{fig:route} summarises the route followed for the piston.

\para{Code.}Every number, table and computed figure in this text is produced by the accompanying
Python code (package \texttt{pistonz}); exact or high-precision arithmetic is used wherever the
algorithms require it.

\section{Continued fractions in a nutshell}\label{sec:nutshell}

A continued fraction is an expression
\begin{equation}\label{eq:cf}
f=b_0+\cfrac{a_1}{b_1+\cfrac{a_2}{b_2+\cfrac{a_3}{b_3+\cdots}}}
\end{equation}
with partial numerators $a_k$ and partial denominators $b_k$, which may depend on a variable $z$.
Stopping after $n$ levels gives the $n$-th approximant $f_n=A_n/B_n$, where
\begin{equation}\label{eq:rec}
A_n=b_nA_{n-1}+a_nA_{n-2},\qquad B_n=b_nB_{n-1}+a_nB_{n-2},
\end{equation}
with $A_{-1}=1$, $A_0=b_0$, $B_{-1}=0$, $B_0=1$. If the $a_k$, $b_k$ are polynomials in $z$, every
approximant is a rational function of $z$.

\para{Evaluation.}The recurrences~\eqref{eq:rec} evaluate top-down, so one can keep adding levels
until $f_n$ stops changing; the \emph{modified Lentz algorithm}~\cite{press} does this with ratios
of successive $A_n$ and $B_n$, which avoids overflow. Bottom-up evaluation is simpler but requires
the depth in advance.

\para{Equivalence.}Multiplying $a_k$, $b_k$ and $a_{k+1}$ by the same non-zero constant leaves the
value unchanged. This is why the same fraction appears with different-looking coefficients in
different books, and why a network may be scaled section by section without changing its
impedance.

\section{From a function to a continued fraction}\label{sec:construct}

\para{From a power series: Viskovatov.}\label{sec:visk}Let $f(z)=c_0+c_1z+c_2z^2+\cdots$. Write
\begin{equation}
f=c_0+\frac{z}{f_1},\qquad f_1=\frac{z}{f-c_0},
\end{equation}
expand $f_1$ again as a power series, and repeat. Each step costs one series division and produces
one partial denominator.

\para{The qd algorithm.}The quotient--difference (qd) algorithm produces the coefficients of the Stieltjes fraction
(S-fraction)
\begin{equation}\label{eq:sfrac}
f(z)=\cfrac{c_0}{1-\cfrac{a_1z}{1-\cfrac{a_2z}{1-\cdots}}}
\end{equation}
directly from the $c_k$. Start with the columns $e_0^{(k)}=0$ and $q_1^{(k)}=c_{k+1}/c_k$ and fill
the table with the \emph{rhombus rules}
\begin{equation}
e_m^{(k)}=q_m^{(k+1)}-q_m^{(k)}+e_{m-1}^{(k+1)},\qquad
q_{m+1}^{(k)}=q_m^{(k+1)}e_m^{(k+1)}/e_m^{(k)}.
\end{equation}
The top row gives $a_1=q_1^{(0)}$, $a_2=e_1^{(0)}$, $a_3=q_2^{(0)}$, $a_4=e_2^{(0)}$, and so on.

For $\ln(1+z)/z$, with $c_k=(-1)^k/(k+1)$, the qd algorithm gives $a_k=-\frac12,-\frac16,-\frac13,-\frac15,\dots$,
i.e.\ $\ln(1+z)/z=1/(1+\frac12z/(1+\frac16z/(1+\cdots)))$ with only positive coefficients, the
signature of a Stieltjes function.

The approximants are Pad\'e approximants of $f$ on a staircase in the Pad\'e table~\cite{baker}:
the $n$-th one agrees with the Taylor series up to order $n$. Both Viskovatov and qd lose accuracy
in floating point, so exact or high-precision arithmetic is advisable; the piston example of
Sec.~\ref{sec:piston} makes this dramatically clear.

If only samples of $f$ are available, Thiele's interpolating continued fraction, built from
inverse divided differences, plays the same role; its accuracy between the nodes depends strongly on
their choice~\cite{cuyt}.

\para{Classical expansions.}Gauss's continued fraction for ratios of hypergeometric functions yields many standard
expansions~\cite{cuyt,wall}, including
\begin{equation}\label{eq:lnclass}
\ln(1+x)=\cfrac{x}{1+\cfrac{x}{2+\cfrac{x}{3+\cfrac{4x}{4+\cfrac{4x}{5+\cfrac{9x}{6+\cdots}}}}}}.
\end{equation}

\section{Beyond the disk of convergence}\label{sec:beyond}

The series $\ln(1+x)=x-x^2/2+x^3/3-\cdots$ converges only for $|x|<1$, because of the
singularity at $x=-1$. The continued fraction~\eqref{eq:lnclass} converges in the whole plane
except on the branch cut $(-\infty,-1]$. Table~\ref{tab:ln3} compares the two at $x=3$.

\begin{table}[htbp]
\caption{$\ln(1+x)$ at $x=3$: the series diverges, the continued fraction~\eqref{eq:lnclass}
converges.}\label{tab:ln3}
\centering\small
\begin{tabular}{rrl}
\toprule
terms & Taylor partial sum & continued fraction\\
\midrule
2 & $-1.5$ & 1.2\\
6 & $-85.7$ & 1.383673\\
10 & $-4311$ & 1.386261\\
12 & $-32494$ & 1.386291\\
\midrule
exact & \multicolumn{2}{c}{$\ln4=1.386294$}\\
\bottomrule
\end{tabular}
\end{table}

The coefficients still encode only local information at 0, so accuracy is best near the expansion
point, and more terms are needed further away. Convergence is guaranteed for well-behaved classes,
notably Stieltjes and hypergeometric-type functions. For general analytic functions, Pad\'e
approximants converge only in measure and may carry spurious pole--zero pairs. A continued fraction
never crosses a genuine branch cut; it approximates a single sheet.

\para{The half-line $[0,\infty)$.}Convergence at every point does not mean uniform accuracy. A
truncated fraction is rational and behaves like $Cx^m$ as $x\to\infty$, so it cannot follow
logarithmic growth: with 20 terms the relative error of~\eqref{eq:lnclass} is $5\times10^{-6}$ at
$x=10$ but 0.37 at $x=10^4$. Four remedies are in common use:
\begin{enumerate}
\item \emph{Two-point Pad\'e}: match expansions at both 0 and $\infty$ (Sec.~\ref{sec:twopoint}).
\item \emph{Factor out the asymptotics}: write $f=(1+x)^\alpha g$ or $f=\ln(1+x)\,g$ with $g$
bounded, and approximate $g$.
\item \emph{Map the range}: substitute $t=x/(1+x)\in[0,1)$ and approximate in $t$ with Thiele
interpolation, Chebyshev nodes or the AAA algorithm~\cite{aaa}.
\item \emph{Split the range}: use the series at 0 for small $x$ and the asymptotic series for
large $x$.
\end{enumerate}
Stieltjes-type functions that vanish at infinity, such as $\ln(1+x)/x$, $e^xE_1(x)$ and Bessel
ratios, have fractions that converge uniformly on $[0,\infty)$.

\section{Two-point Pad\'e approximants}\label{sec:twopoint}

Suppose $f$ has expansions at both ends,
\begin{equation}
f\sim\sum_{j\ge0}c_jx^j\ (x\to0),\qquad f\sim\sum_{j\ge0}d_jx^{-j}\ (x\to\infty).
\end{equation}
Seek $R=P/Q$ with $\deg P=\deg Q=n$. After normalising $Q(0)=1$ there are $2n+1$ free
coefficients. Impose $k$ conditions at 0 and $2n+1-k$ at $\infty$:
\begin{equation}\label{eq:tp}
Qf-P=O(x^k)\ (x\to0),\qquad Qf-P=O(x^{k-n-1})\ (x\to\infty).
\end{equation}
Both sets are linear in the coefficients of $P$ and $Q$. Raising the order one step at a time,
alternating between the two ends, gives a staircase of approximants linked by a three-term
recurrence: they are the convergents of an \emph{M-fraction}, whose coefficients follow from the
$c_j$, $d_j$ by the two-point qd algorithm~\cite{mccabe}.

\begin{figure}[htbp]
\centering
\includegraphics[width=0.62\textwidth]{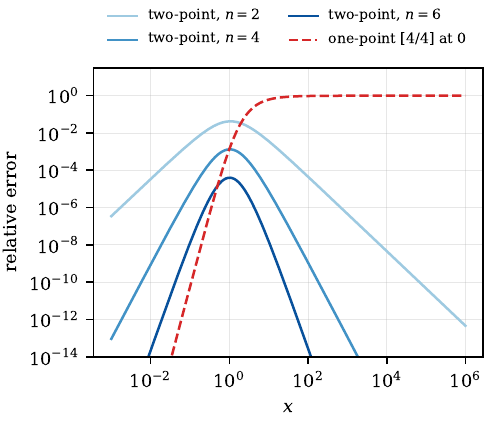}
\caption{Relative error of approximants to $\arctan x$. The two-point errors vanish at both ends,
where the expansions are matched, and peak in between. The one-point approximant is accurate only
for small $x$.}\label{fig:arctan}
\end{figure}

\para{Example: $\arctan x$.}At 0, $\arctan x=x-x^3/3+\cdots$; at $\infty$,
$\arctan x=\pi/2-1/x+1/(3x^3)-\cdots$. For $n=2$, $k=3$:
\begin{equation}
\arctan x\approx\frac{x\,(2\pi x+\pi^2-4)}{4x^2+2\pi x+\pi^2-4},
\end{equation}
with unit slope at 0 and limit $\pi/2$ at $\infty$. Figure~\ref{fig:arctan} compares the two-point approximants ($k=n+1$) with the one-point $[n/n]$ Pad\'e
approximant at 0 over $10^{-3}\le x\le10^6$: the maximum relative error of the two-point approximant is
$4.2\times10^{-2}$, $1.3\times10^{-3}$ and $4.0\times10^{-5}$ for $n=2$, 4 and 6, whereas the one-point
approximant is 100\% off at large $x$.

The one-point approximant tends to 0 or $\infty$ at large $x$ (for odd $n$ the $[n/n]$ denominator
degenerates and the approximant diverges). The two-point error falls steadily with $n$, uniformly
over the half-line. In practice the expansion at $\infty$ must be in integer powers of $1/x$
(remove $\ln x$ or $\sqrt x$ terms first, e.g.\ with $x=t^2$), a split $k\approx n$ usually gives
the best uniform error, and the linear system should be solved in exact or high-precision
arithmetic.

\section{Ladder networks}\label{sec:networks}

\subsection{A ladder is a continued fraction}
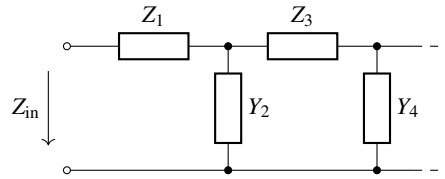
\begin{figure}[htbp]
\centering
\begin{circuitikz}[scale=0.82, transform shape]
\draw (0,2) to[short,o-] (0.4,2) to[generic,l=$Z_1$] (2.4,2) -- (2.8,2) to[generic,l=$Z_3$] (4.8,2) -- (5.6,2);
\draw[dashed] (5.6,2) -- (6.2,2);
\draw (0,0) to[short,o-] (5.6,0);
\draw[dashed] (5.6,0) -- (6.2,0);
\draw (2.6,2) to[generic,l=$Y_2$,*-*] (2.6,0);
\draw (5.0,2) to[generic,l=$Y_4$,*-*] (5.0,0);
\draw[->] (-0.3,1.6) -- (-0.3,0.4) node[midway,left]{$Z_\mathrm{in}$};
\end{circuitikz}
\caption{A ladder network. Its input impedance is the continued fraction~\eqref{eq:ladder}.}
\label{fig:ladder}
\end{figure}

Consider the ladder of Figure~\ref{fig:ladder}, with series impedances $Z_1,Z_3,\dots$ and shunt
admittances $Y_2,Y_4,\dots$. Looking into the input one sees $Z_1$ in series with whatever lies
behind it, that is $Y_2$ in parallel with the rest, and so on. Hence
\begin{equation}\label{eq:ladder}
Z_\mathrm{in}=Z_1+\cfrac{1}{Y_2+\cfrac{1}{Z_3+\cfrac{1}{Y_4+\cdots}}}.
\end{equation}
With $Z_k=sL_k$ and $Y_k=sC_k$ every partial denominator is one physical component. Reading off a
continued fraction is therefore the same as drawing a circuit.

The functions that can occur are \emph{positive real}: $\Re Z(s)\ge0$ for $\Re s\ge0$. For lossless
($LC$) networks $Z$ is an odd rational function whose poles and zeros alternate on the imaginary
axis (Foster's reactance theorem~\cite{foster}).

\subsection{Cauer synthesis: worked example}
Take the reactance function
\begin{equation}\label{eq:react}
Z(s)=\frac{s^4+4s^2+3}{s^3+2s}.
\end{equation}

\para{Cauer I (expansion at $s=\infty$).}Long division from the highest powers gives
$Z=s+(2s^2+3)/(s^3+2s)$. Inverting the remainder and dividing again, and repeating, yields
\begin{equation}
Z(s)=s+\cfrac{1}{\frac s2+\cfrac{1}{4s+\cfrac{1}{\frac s6}}},
\end{equation}
that is, series $L_1=1$\,H, shunt $C_2=\frac12$\,F, series $L_3=4$\,H, shunt $C_4=\frac16$\,F: a
low-pass ladder.

\begin{figure}[htbp]
\centering
\begin{circuitikz}[scale=0.72, transform shape]
\node at (2.3,3.25) {\textbf{Cauer I}};
\draw (0,2) to[short,o-] (0.2,2) to[L,l=1\,H] (1.8,2) -- (2.2,2) to[L,l=4\,H] (3.8,2) -- (4.6,2);
\draw (0,0) to[short,o-] (4.6,0);
\draw (2.0,2) to[C,l_=$\frac12$\,F,*-*] (2.0,0);
\draw (4.6,2) to[C,l_=$\frac16$\,F] (4.6,0);
\begin{scope}[xshift=5.6cm]
\node at (2.3,3.25) {\textbf{Cauer II}};
\draw (0,2) to[short,o-] (0.2,2) to[C,l=$\frac23$\,F] (1.8,2) -- (2.2,2) to[C,l=$\frac2{25}$\,F] (3.8,2) -- (4.6,2);
\draw (0,0) to[short,o-] (4.6,0);
\draw (2.0,2) to[L,l_=$\frac54$\,H,*-*] (2.0,0);
\draw (4.6,2) to[L,l_=5\,H] (4.6,0);
\end{scope}
\end{circuitikz}
\caption{Two ladder realisations of $Z(s)=(s^4+4s^2+3)/(s^3+2s)$: expansion at $s=\infty$ (left)
and at $s=0$ (right).}\label{fig:cauer}
\end{figure}
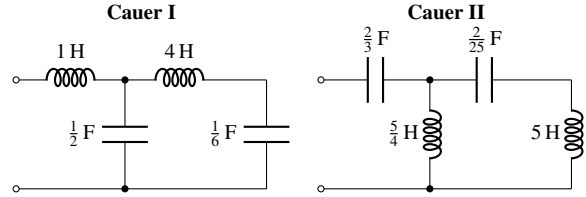

\para{Cauer II (expansion at $s=0$).}Dividing from the lowest powers instead gives
\begin{equation}
Z(s)=\frac{3}{2s}+\cfrac{1}{\frac{4}{5s}+\cfrac{1}{\frac{25}{2s}+\cfrac{1}{\frac{1}{5s}}}},
\end{equation}
that is, series $C_1=\frac23$\,F, shunt $L_2=\frac54$\,H, series $C_3=\frac2{25}$\,F, shunt
$L_4=5$\,H: a high-pass ladder (Figure~\ref{fig:cauer}). Both reproduce~\eqref{eq:react} exactly.

Two points deserve emphasis. First, Cauer I is a one-point continued-fraction expansion at
$s=\infty$ and Cauer II one at $s=0$; the network topology is a direct picture of where the
expansion is made. Second, the network is realisable precisely when all quotients are positive.
The same test, applied to the ratio of the even and odd parts of a polynomial, is the Routh--Hurwitz
stability criterion: the Routh array is a continued-fraction expansion in disguise.

\subsection{RC networks are Stieltjes functions}
The driving-point impedance of any $RC$ network has the form
\begin{equation}
Z(s)=k_\infty+\frac{k_0}{s}+\sum_i\frac{k_i}{s+\sigma_i},\qquad k\ge0,\ \sigma_i>0,
\end{equation}
with poles and zeros alternating on the negative real axis. This is a Stieltjes function, the class
for which the theory of Secs.~\ref{sec:construct}--\ref{sec:twopoint} is strongest: its S-fraction
has positive coefficients, its Pad\'e approximants converge with poles on the negative axis, and
every truncation is again realisable as a positive $RC$ ladder. Truncating the continued fraction is
thus identical to building a smaller physical circuit. The piston impedance is \emph{not} of this
kind (its poles are complex), which is what makes it interesting.

\subsection{Two-point Pad\'e network reduction}\label{sec:rcline}
\para{Problem.}A uniform $RC$ line (a resistive track over a ground plane, or a diffusion
electrode) is modelled by $N=20$ sections: shunt $C/N$ first, then series $R/N$, ending in a short,
with $R=C=1$. Its impedance satisfies $Z(0)=1$ and $Z(s)\approx20/s-8000/s^2$ as $s\to\infty$. We
want a model with only three capacitors and three resistors ($n=3$, six free coefficients). We match
$k$ coefficients at $s=0$ and $6-k$ at $s=\infty$, solve the linear system exactly, and synthesise
the result as a Cauer I ladder.

\begin{table}[htbp]
\caption{Three-section $RC$ models of a 20-section line, matching $k$ coefficients at DC and $6-k$
at infinity. Last column: maximum relative error of $Z(\mathrm{j}\omega)$ for
$10^{-2}\le\omega\le10^5$.}\label{tab:rc}
\centering\small
\setlength{\tabcolsep}{3.6pt}
\begin{tabular}{rccccccc}
\toprule
$k$ & $C_1$ & $R_1$ & $C_2$ & $R_2$ & $C_3$ & $R_3$ & max err\\
\midrule
0 & .050 & .050 & .050 & .050 & .050 & .050 & 0.85\\
1 & .050 & .050 & .050 & .050 & .050 & .900 & 0.95\\
2 & .050 & .050 & .050 & .050 & .325 & .900 & 0.30\\
3 & .050 & .050 & .050 & .180 & .445 & .770 & 0.25\\
4 & .050 & .050 & .122 & .277 & .439 & .673 & \textbf{0.17}\\
5 & .050 & .093 & .193 & .303 & .410 & .604 & 0.21\\
6 & .077 & .145 & .225 & .300 & .381 & .555 & 0.35\\
\bottomrule
\end{tabular}
\end{table}

Table~\ref{tab:rc} reveals the structure clearly.
\begin{itemize}
\item \emph{Matching $m=6-k$ coefficients at infinity reproduces exactly the first $m$ elements of
the original ladder} (the entries 0.050). This is no accident: Cauer I is the expansion at
$\infty$, so the high-frequency moments are the input-side elements.
\item $k=0$ is simply the original ladder cut off after six elements. Its DC resistance is 0.15
instead of 1, so it is useless at low frequency: naive truncation is a one-point Pad\'e
approximant at $\infty$.
\item $k=6$ is the one-point Pad\'e approximant at DC. It gets $R_\mathrm{dc}=1$ but the wrong input
capacitance (0.077 instead of 0.050), so it is 35\% off at high frequency.
\item Every $k\ge1$ gives total resistance exactly 1, and every element value is positive, as
expected for a Stieltjes function. The best uniform error occurs for a split $k=4$, close to the
middle: its first two elements are copied from the original ladder, the others are fixed by the four
DC moments.
\end{itemize}

\subsection{Wider practice}
The same moment-matching idea underlies model-order reduction in circuit simulation: asymptotic
waveform evaluation (AWE)~\cite{awe}, Pad\'e-via-Lanczos (PVL)~\cite{pvl} and PRIMA~\cite{prima}
match moments at one or several expansion points, PRIMA additionally guaranteeing passivity. Vector
fitting~\cite{gustavsen,sanathanan} matches measured data at chosen frequencies, the network
analogue of Thiele interpolation. Distributed elements such as the Warburg impedance
($\propto s^{-1/2}$), transmission lines and skin-effect losses in loudspeaker voice coils have
non-integer behaviour at infinity; as in Sec.~\ref{sec:twopoint}, a change of variable or factoring
out the $\sqrt s$ behaviour comes first.

\section{Case study 1: the baffled piston}\label{sec:piston}

\subsection{The function and its expansions}\label{sec:func}
\para{Normalisation.}Impedance is normalised by $\rc$ and time by $a/c$, so that $s=\mathrm{j}ka$.
Normalised masses convert to physical ones by multiplication with $\rho_0Sa=\rho_0\pi a^3$,
compliances by multiplication with $a/(\rho_0c^2S)$, and resistances by multiplication with $\rc$.

\para{Laplace-domain form.}With $J_1(-\mathrm{j}x)=-\mathrm{j}I_1(x)$ and the modified Struve
function $\mathbf{L}_1$, Eq.~\eqref{eq:Z} becomes
\begin{equation}\label{eq:Zs}
Z(s)=1-\frac{I_1(2s)-\mathbf{L}_1(2s)}{s},
\end{equation}
an \emph{entire} function of $s$: it has no poles at all, only an essential singularity at
infinity.

\para{Expansion at $s=0$.}The series converges everywhere~\cite[Eqs.~9.1.10, 12.1.3]{as}:
\begin{equation}\label{eq:taylor}
\begin{gathered}
Z(s)=\sum_{n\ge1}c_ns^n=\frac{8}{3\pi}s-\frac12s^2+\frac{32}{45\pi}s^3-\frac1{12}s^4+\cdots,\\
c_{2m+1}=\frac{1}{\Gamma(m+\frac32)\Gamma(m+\frac52)},\qquad c_{2m}=\frac{-1}{m!\,(m+1)!}.
\end{gathered}
\end{equation}
The first term is the radiation mass $M_0=8/3\pi$ (physically $8\rho_0a^3/3$); the second gives
the radiation resistance $\frac12(ka)^2$.

\para{Expansion at $s=\infty$.}For $|s|\to\infty$ with $|\arg s|<\pi/2$, the asymptotic expansion
of $I_1-\mathbf{L}_1$ gives the divergent series
\begin{equation}\label{eq:asym}
\begin{aligned}
Z(s)&\sim1+\frac1\pi\sum_{k\ge0}\frac{(-1)^{k+1}\Gamma(k+\frac12)}{\Gamma(\frac32-k)}\,s^{-2k-1}\\
&=1-\frac{2}{\pi s}+\frac{1}{2\pi s^3}+\frac{3}{8\pi s^5}+\frac{45}{32\pi s^7}+\cdots,
\end{aligned}
\end{equation}
in integer powers of $1/s$, which is exactly what a two-point Pad\'e approximant needs. On the
frequency axis itself, however, oscillatory terms of the type $J_1(2ka)/ka$ add to~\eqref{eq:asym}.
They decay only like $(ka)^{-3/2}$ and are the ripples visible in $\Re Z$ and $\Im Z$. No rational
function with poles away from the axis can reproduce them, so they set an error floor for every
rational method.

\para{Delay representation.}The Poisson integrals of $J_1$ and $\Ho$~\cite[Eqs.~9.1.20, 12.1.7]{as} give
\begin{equation}\label{eq:delay}
Z(s)=\frac4\pi\int_0^{\pi/2}\sin^2\alpha\,\bigl(1-e^{-2s\cos\alpha}\bigr)\,\mathrm{d}\alpha,
\end{equation}
a continuum of delays, one for every chord $2a\cos\alpha$ of the piston. It yields the impulse
response $\delta(t)-(2/\pi)\sqrt{1-t^2/4}$ for $0<t<2$ (in units of $a/c$) and zero afterwards.

\subsection{One-point continued fraction and its even contraction}\label{sec:cfpiston}
\begin{figure}[htbp]
\centering
\includegraphics[width=0.62\textwidth]{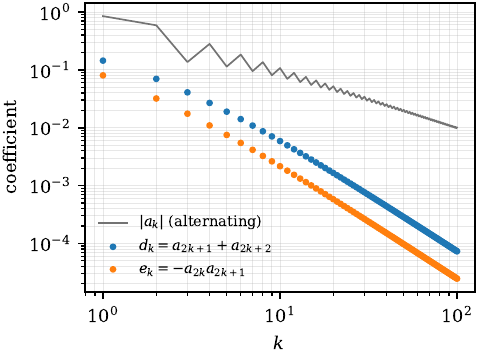}
\caption{C-fraction coefficients $|a_k|$ (alternating in sign from $k=3$) and the coefficients
$d_k$, $e_k$ of the even contraction, all positive.}\label{fig:coef}
\end{figure}

The series~\eqref{eq:taylor} is expanded into a C-fraction~\cite{wall,baker}
\begin{equation}\label{eq:cfrac}
Z(s)=\cfrac{a_1s}{1+\cfrac{a_2s}{1+\cfrac{a_3s}{1+\cdots}}},
\end{equation}
by the Viskovatov algorithm of Sec.~\ref{sec:visk}, in exact rational--$\pi$ arithmetic or, as
here, 1000-digit floating point. The first coefficients are $a_1=8/3\pi$, $a_2=3\pi/16$,
$a_3=-0.13634$, $a_4=0.28075$, $a_5=-0.11394$. From $a_3$ onwards the signs alternate
(Figure~\ref{fig:coef}): $Z$ is not a Stieltjes function, so~\eqref{eq:cfrac} cannot itself be
realised as an $RL$ or $RC$ ladder. That is to be expected, since such ladders have only real poles
and cannot produce the overshoot of $\Re Z$.

The \emph{even contraction} of~\eqref{eq:cfrac}~\cite{wall} combines consecutive levels into a
J-fraction:
\begin{equation}\label{eq:jfrac}
Z(s)=\cfrac{a_1s}{1+a_2s+\cfrac{e_1s^2}{1+d_1s+\cfrac{e_2s^2}{1+d_2s+\cdots}}},
\end{equation}
with
\begin{equation}\label{eq:de}
d_k=a_{2k+1}+a_{2k+2},\qquad e_k=-a_{2k}\,a_{2k+1}.
\end{equation}
Truncating~\eqref{eq:jfrac} after $d_n$ gives the same rational function as truncating~\eqref{eq:cfrac}
after $a_{2n+2}$: the diagonal Pad\'e approximant that matches the first $2n+2$ Taylor coefficients.
All computed $d_k$ and $e_k$ are positive (Figure~\ref{fig:coef}), decaying smoothly as $k^{-2}$.
This positivity is the key to everything that follows.

\subsection{Ladder realisation}\label{sec:lad}
\begin{figure}[htbp]
\centering
\begin{circuitikz}[scale=0.7, transform shape]
\draw (0,2) to[short,o-] (0.8,2) -- (2.2,2) to[C,l=$C_1$] (3.6,2) to[R,l=$R_1$] (5.0,2) -- (6.6,2) to[C,l=$C_3$] (8.0,2) to[R,l=$R_3$] (9.4,2);
\draw[dashed] (9.4,2) -- (10.0,2);
\draw (0,0) to[short,o-] (9.4,0);
\draw[dashed] (9.4,0) -- (10.0,0);
\draw (0.8,2) to[L,l_=$M_0$,*-*] (0.8,0);
\draw (2.2,2) to[R,l=$1/G_0$,*-*] (2.2,0);
\draw (5.0,2) to[L,l_=$M_2$,*-*] (5.0,0);
\draw (6.4,2) to[R,l=$1/G_2$,*-*] (6.4,0);
\draw[->] (-0.3,1.6) -- (-0.3,0.4) node[midway,left]{$Z$};
\end{circuitikz}
\caption{The piston ladder in the mechanical-impedance analogy (inductor${}={}$mass,
capacitor${}={}$compliance, resistor${}={}$mechanical resistance). Each new order appends a series
C--R section or a shunt M$\parallel$R section at the open end; nothing to its left changes.}
\label{fig:pladder}
\end{figure}
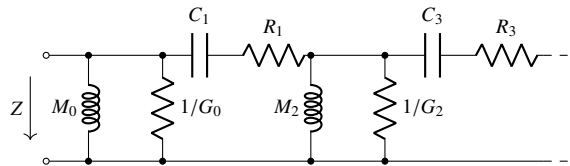

Writing $Y=1/Z$, \eqref{eq:jfrac} reads
\begin{equation}\label{eq:Y}
Y(s)=\frac{1}{a_1s}+\frac{a_2}{a_1}+\frac{e_1}{a_1}\,\frac{s}{1+d_1s+e_2s^2/(\dots)}.
\end{equation}
The first two terms are a shunt mass $M_0=a_1$ and a shunt conductance $G_0=a_2/a_1=9\pi^2/128$. The
third is the admittance of an impedance $\kappa_1^{-1}(1+d_1s+\dots)/s$: a series compliance
$\kappa_1$, a series resistance $d_1/\kappa_1$ and a remainder of the same form, with the roles of
impedance and admittance exchanged. Repeating the argument gives the ladder of Figure~\ref{fig:pladder},
in which section $j=1,2,\dots,n$ consists of
\begin{equation}\label{eq:elements}
\begin{aligned}
j\text{ odd (series):}&\quad C_j=\kappa_j,\quad R_j=d_j/\kappa_j,\\
j\text{ even (shunt):}&\quad M_j=\kappa_j,\quad G_j=d_j/\kappa_j,
\end{aligned}
\end{equation}
with $\kappa_1=e_1/a_1$ and $\kappa_{j+1}=e_{j+1}/\kappa_j$. Throughout, $N$ denotes the total number of
physical elements (masses, compliances and resistances, including the radiation mass $M_0$ and the
resistance $1/G_0$ of section 0) and $n$ the number of further sections, so that $N=2+2n$. The
ladder with $N$ elements reproduces the Taylor coefficients of $s^1,\dots,s^N$ in~\eqref{eq:taylor}
($Z(0)=0$ holds automatically): each element adds one matched coefficient, as was verified
numerically up to $N=14$. It is the mixed Cauer form of Sec.~\ref{sec:networks}, with one compliance or
mass and one resistance per level.

\para{Passivity.}If all $d_k,e_k>0$, all $\kappa_j$ and all element values are positive. Every
truncation is then the driving-point impedance of a passive network, hence positive real, with all
poles in the open left half-plane, and stable in any state-space realisation.

\para{Nesting.}Raising the order from $n$ to $n+1$ uses $d_{n+1},e_{n+1}$ only: one section is
appended at the open end and all existing values are unchanged.

\para{First elements.}In closed form,
\begin{equation}\label{eq:first}
\begin{aligned}
M_0&=\frac{8}{3\pi}, & 1/G_0&=\frac{128}{9\pi^2},\\
C_1&=\frac{27\pi^3}{2048}-\frac{\pi}{10}, & R_1&=1.52624.
\end{aligned}
\end{equation}
Table~\ref{tab:elements} lists the first 26 elements. $1/G_j$ grows by almost exactly 3 per shunt
section, while $C_j$, $M_j$ and $R_j$ decrease roughly as $1/j$.

\begin{table}[htbp]
\caption{Ladder element values (normalised). Section 0 is the radiation mass and resistance; odd
sections are series, even sections shunt. A ladder of $N=2+2n$ elements uses sections $0\dots n$.}
\label{tab:elements}
\centering\small
\begin{tabular}{rlll}
\toprule
$j$ & type & mass / compliance & resistance\\
\midrule
0 & shunt & $M=0.848826$ & $1/G=1.441012$\\
1 & series & $C=0.094615$ & $R=1.526242$\\
2 & shunt & $M=0.338096$ & $1/G=4.837913$\\
3 & series & $C=0.051634$ & $R=0.791043$\\
4 & shunt & $M=0.212060$ & $1/G=7.912781$\\
5 & series & $C=0.035362$ & $R=0.535458$\\
6 & shunt & $M=0.154216$ & $1/G=10.946181$\\
7 & series & $C=0.026863$ & $R=0.405457$\\
8 & shunt & $M=0.121099$ & $1/G=13.964285$\\
9 & series & $C=0.021650$ & $R=0.326469$\\
10 & shunt & $M=0.099669$ & $1/G=16.974925$\\
11 & series & $C=0.018129$ & $R=0.273320$\\
12 & shunt & $M=0.084674$ & $1/G=19.981383$\\
\bottomrule
\end{tabular}
\end{table}

\subsection{Positivity: a computer-assisted proof}\label{sec:pos}
\para{Theorem.}$d_k>0$ and $e_k>0$ for all $1\le k\le100$. Hence every ladder of up to 202 elements
built from~\eqref{eq:elements} has only positive elements and is passive.

\para{Proof (computer-assisted).}The Viskovatov recursion and the contraction are evaluated in
interval arithmetic with outward rounding at 4000-bit precision. The inputs are enclosed rigorously:
factorials are exact integers, $\pi$ is an interval, and
$\Gamma(n+\frac32)=\sqrt\pi\,(2n+2)!/(4^{n+1}(n+1)!)$. No division involves an interval that contains
zero, so every coefficient provably exists. For every $k\le100$ the lower endpoints of the enclosures
of $d_k$ and $e_k$ are positive; the enclosures remain tight to at least 800 digits. The computation
takes about one minute.\hfill$\Box$

\para{Conditioning.}The recursion is severely ill-conditioned. Rounding the factorials to double
precision (relative error $\approx10^{-16}$) alone produces a spurious negative $d_{16}$. With exact
inputs it is positive. This is the floating-point warning of Sec.~\ref{sec:construct} in its sharpest
form.

\para{Asymptotics.}Numerically, with corrections decaying like $1/k$,
\begin{equation}\label{eq:kasym}
k^2e_k\to\tfrac14,\qquad k^2d_k\to\tfrac34\qquad(k\to\infty);
\end{equation}
at $k=98$ the values are 0.2462 and 0.7303. The ratio $d_k/e_k\to3$ explains why $1/G_j$ grows by 3
per shunt section. For the single delay $1-e^{-2s}$, the end point of the delay
continuum~\eqref{eq:delay}, the same construction gives $d_k=0$ and $e_k=1/(4k^2-1)$, so the limit
$\frac14$ appears to be set by the longest delay.

\para{All $k$.}Positivity for all $k$ remains a conjecture. The usual route, a Stieltjes-type
representation by a positive measure, does not apply directly: in the equivalent Jacobi form the
off-diagonal entries are positive but the diagonal is imaginary (the dissipative case). Explicit
remainder bounds in~\eqref{eq:kasym}, combined with the theorem, would complete the proof.

\subsection{Accuracy of the nested ladder}\label{sec:res}
\begin{figure}[htbp]
\centering
\includegraphics[width=0.95\textwidth]{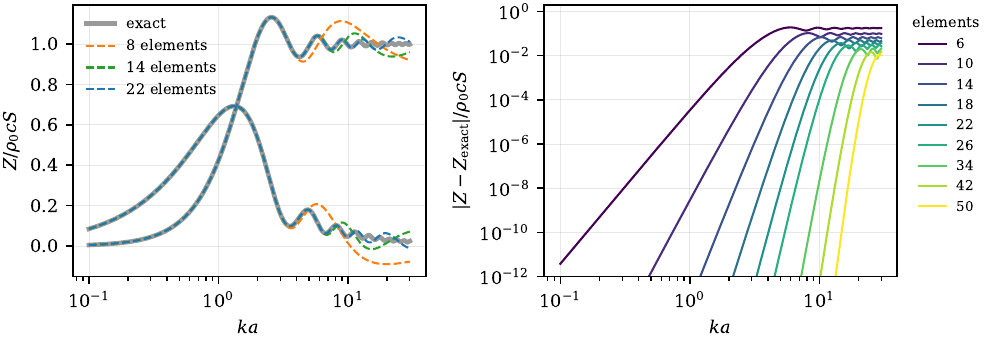}
\caption{Left: exact impedance (grey) and nested ladders of 8, 14 and 22 elements (dashed). Right:
absolute error for ladders of 6 to 50 elements.}\label{fig:Z}
\end{figure}

Unless stated otherwise, errors are absolute, $|Z-Z_\mathrm{exact}|/\rc$; relative errors
$|Z/Z_\mathrm{exact}-1|$ appear only in Table~\ref{tab:methods}, where they weigh the low frequencies
at which $|Z|\to0$.
Figure~\ref{fig:Z} compares ladders of 8, 14 and 22 elements with~\eqref{eq:Z}, and
Table~\ref{tab:error} gives the maximum error. As expected for a Pad\'e approximant at $s=0$, the
accuracy is extreme at low $ka$ and degrades beyond a cut-off that moves upwards as elements are
added. Already 14 elements give $4\times10^{-4}$ up to $ka=5$, and 26 elements give
$1.1\times10^{-4}$ up to $ka=10$. From 50 elements onwards the ladder is exact to double precision
for $ka\le10$. The Taylor series, by contrast, although convergent for every $s$, has terms that
grow enormously before they decay: truncated after $s^9$ its relative error over
$0.005\le ka\le50$ is $10^{12}$. Convergence in principle is not accuracy in practice.

\begin{table}[htbp]
\caption{Nested ladder: maximum error $|Z-Z_\mathrm{exact}|/\rc$ over $ka\le\kappa$, and the
largest $ka$ up to which the error stays below $10^{-2}$.}\label{tab:error}
\centering\small
\setlength{\tabcolsep}{4.5pt}
\begin{tabular}{rccccc}
\toprule
$N$ & $\kappa=2$ & $\kappa=5$ & $\kappa=10$ & $\kappa=20$ & $ka_{1\%}$\\
\midrule
4 & 3.7e-02 & 2.8e-01 & 2.8e-01 & 2.8e-01 & 1.5\\
6 & 3.0e-03 & 1.6e-01 & 1.9e-01 & 1.9e-01 & 2.5\\
8 & 1.4e-04 & 6.2e-02 & 1.4e-01 & 1.4e-01 & 3.5\\
10 & 4.6e-06 & 1.7e-02 & 1.0e-01 & 1.0e-01 & 4.6\\
12 & 1.0e-07 & 3.2e-03 & 8.2e-02 & 8.3e-02 & 5.7\\
14 & 1.8e-09 & 4.3e-04 & 6.6e-02 & 6.8e-02 & 6.9\\
18 & 2.6e-13 & 3.2e-06 & 2.1e-02 & 4.9e-02 & 9.1\\
22 & $<10^{-15}$ & 9.8e-09 & 2.4e-03 & 3.8e-02 & 11.4\\
26 & $<10^{-15}$ & 1.5e-11 & 1.1e-04 & 2.9e-02 & 13.7\\
34 & $<10^{-15}$ & $<10^{-15}$ & 2.2e-08 & 1.8e-02 & 18.3\\
42 & $<10^{-15}$ & $<10^{-15}$ & 5.4e-13 & 1.5e-03 & 23.1\\
50 & $<10^{-15}$ & $<10^{-15}$ & $<10^{-15}$ & 9.5e-06 & 28.2\\
\bottomrule
\end{tabular}
\end{table}

\para{Bandwidth.}The range of validity grows linearly with the number of elements
(the curves of Figure~\ref{fig:Z}, right). The error stays below 1\% up to $ka\approx0.58N-1.2$, and below 0.1\%
up to $ka\approx0.50N-1.6$ (least-squares fits for $4\le N\le50$). Each additional pair of elements
extends the valid range by about 1.2 units of $ka$.

\begin{figure}[htbp]
\centering
\includegraphics[width=0.62\textwidth]{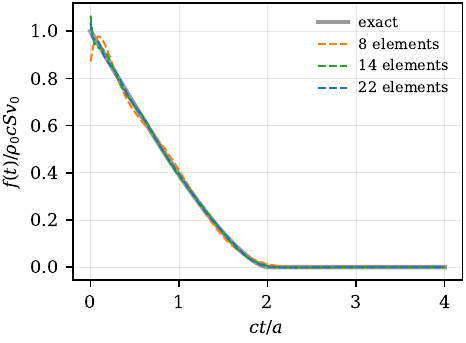}
\caption{Force for a unit velocity step: exact~\eqref{eq:step} and ladders of 8, 14 and 22
elements.}\label{fig:step}
\end{figure}

\para{Time domain.}The force for a unit velocity step is
\begin{equation}\label{eq:step}
\frac{f(t)}{\rc v_0}=1-\frac{2}{\pi}\Big[\tau\sqrt{1-\tau^2}+\arcsin\tau\Big],\quad
\tau=\frac{ct}{2a}\le1,
\end{equation}
and $f=0$ for $t>2a/c$. The ladders reproduce this closely (Figure~\ref{fig:step}); their step
responses are computed from the poles and residues in high precision, since floating-point
polynomial roots become unreliable above about 20 elements. The maximum deviation for $t>0.1a/c$ is
0.048, 0.018 and 0.008 for 8, 14 and 22 elements. The finite support $t\le2a/c$, the diameter
travel time, is closely approximated, although a rational network cannot reproduce it exactly. The initial value $Z(\infty)$ is not prescribed by a one-point expansion
and oscillates around the exact value 1 (0.87, 1.07 and 1.04).

\subsection{Low order: two-point Pad\'e and a one-parameter family}\label{sec:family}
At low order the one-point ladder is poor at high frequency (0.284 at 4 elements over
$0.1\le ka\le50$), because $Z(\infty)$ is not enforced. Two-point Pad\'e approximants
(Sec.~\ref{sec:twopoint}) built from~\eqref{eq:taylor} and~\eqref{eq:asym} repair this.

\para{Order 1.}With $n=1$, $k=2$ the conditions are $Z(0)=0$, $Z'(0)=8/3\pi$ and $Z(\infty)=1$:
\begin{equation}
Z_1(s)=\frac{M_0s}{1+M_0s},\qquad M_0=\frac{8}{3\pi},
\end{equation}
a resistance $\rc$ in parallel with the radiation mass, the classical simple circuit of a baffled
piston~\cite{beranek}, here obtained as the lowest-order two-point Pad\'e approximant.

\para{Order 2.}With $n=2$, $k=3$ (three conditions at 0, two at $\infty$) the approximant is, in
closed form,
\begin{equation}
Z_2(s)=\frac{M_0s+p\,s^2}{1+q\,s+p\,s^2},\quad p=\frac{128-9\pi^2}{18\pi^2-96},\quad
q=\frac{5\pi}{3\pi^2-16}.
\end{equation}
A partial-fraction expansion of $1/Z_2$ gives a mass $M_0$, a resistance $R_a$, and a resistance
$R_b$ in series with a compliance $C$, all in parallel, with
\begin{equation}\label{eq:order2}
\begin{gathered}
R_a=\frac{128}{9\pi^2}=1.441,\qquad R_b=\frac{128}{128-9\pi^2}=3.268,\\
C=\frac{\pi(128-9\pi^2)^2}{2048(3\pi^2-16)}=0.1730.
\end{gathered}
\end{equation}
$R_a$ reproduces the low-frequency radiation resistance $\frac12(ka)^2$ through the parallel
$R_a$--$M_0$ combination, and $R_a\parallel R_b=1$ fixes the high-frequency limit.

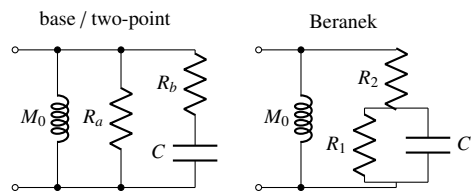
\begin{figure}[htbp]
\centering
\begin{circuitikz}[scale=0.7, transform shape]
\draw (0,2.6) to[short,o-] (3.4,2.6);
\draw (0,0) to[short,o-] (3.4,0);
\draw (0.8,2.6) to[L,l_=$M_0$,*-*] (0.8,0);
\draw (2.0,2.6) to[R,l_=$R_a$,*-*] (2.0,0);
\draw (3.4,2.6) to[R,l_=$R_b$] (3.4,1.3) to[C,l_=$C$] (3.4,0);
\node at (1.7,3.15) {base / two-point};
\begin{scope}[xshift=4.6cm]
\draw (0,2.6) to[short,o-] (2.6,2.6);
\draw (0,0) to[short,o-] (2.6,0);
\draw (0.8,2.6) to[L,l_=$M_0$,*-*] (0.8,0);
\draw (2.6,2.6) to[R,l_=$R_2$] (2.6,1.5);
\draw (2.0,1.5) -- (3.2,1.5);
\draw (2.0,1.5) to[R,l_=$R_1$] (2.0,0.1);
\draw (3.2,1.5) to[C,l=$C$] (3.2,0.1);
\draw (2.0,0.1) -- (3.2,0.1); \draw (2.6,0.1) -- (2.6,0);
\node at (1.6,3.15) {Beranek};
\end{scope}
\end{circuitikz}
\caption{The four-element family. Left: base network and order-two two-point Pad\'e approximant,
Eq.~\eqref{eq:order2}. Right: Beranek's network. Both are $M_0$ in parallel with the branch
admittance~\eqref{eq:fam}.}\label{fig:famcirc}
\end{figure}

\begin{figure}[htbp]
\centering
\includegraphics[width=0.95\textwidth]{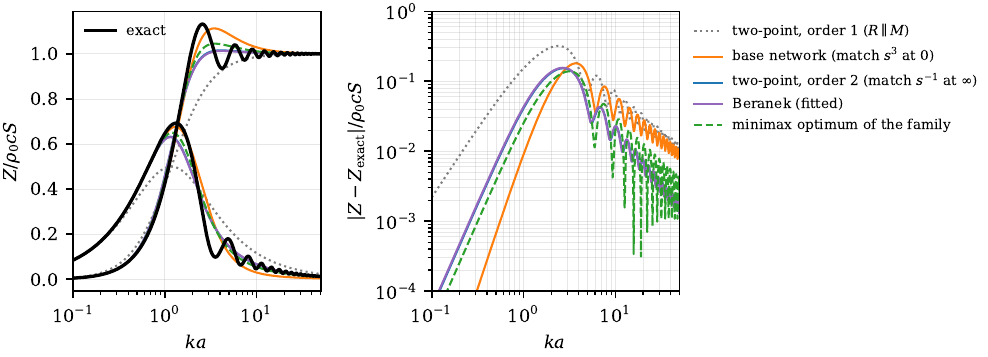}
\caption{The four-element family. Left: exact impedance (black) and the members. Right: their
errors. The two-point approximant (blue) and Beranek's network (purple) almost coincide. The
order-one two-point approximant ($R\parallel M$, dotted) is shown for comparison.}\label{fig:family}
\end{figure}

\para{The family.}This is the same topology as the \emph{base network}, the four-element nested
ladder whose last resistor is set by $Z(\infty)=1$ (variant (a) of Sec.~\ref{sec:term} with $N=4$;
left in Figure~\ref{fig:famcirc}), and Beranek's
network~\cite[Fig.~4.37(a), Eqs.~(4.147)--(4.151), pp.~200--201]{beranekmellow2019} (right) is
equivalent to it: a mass in parallel with $R_2+(R_1\parallel C)$, normalised $R_2=1$,
$R_1+R_2=128/9\pi^2$ and $C=1.89$, the latter chosen for the best average fit. All of them consist of
$M_0$ in parallel with a first-order branch whose admittance is $G_0$ at $s=0$ and 1 at
$s=\infty$,
\begin{equation}\label{eq:fam}
Y_\mathrm{b}(s)=\frac{1+s\tau}{128/9\pi^2+s\tau},
\end{equation}
and they differ only in the time constant $\tau$. With the three exact conditions $M_0$,
$\frac12(ka)^2$ and $Z(\infty)=1$ fixed, one parameter remains, and each classical network is one way
of choosing it (Figure~\ref{fig:family}; broadband errors over $0.1\le ka\le50$ in brackets):
\begin{itemize}
\item the base network matches the third Taylor coefficient at $s=0$ ($\tau=0.445$; 0.181);
\item the two-point approximant matches the $1/s$ term at $s=\infty$ ($\tau=0.814$; 0.153);
\item Beranek fits it to the curves ($\tau=0.834$; 0.156);
\item the minimax optimum over $0.1\le ka\le50$ has $\tau=\tauOpt$ (0.139).
\end{itemize}
Beranek's fitted value $C=1.89$ thus corresponds, to about 2\%, to the two-point value
$C=\tau/R_1=1.85$ in his topology: \emph{Beranek's network is, to graphical accuracy, the order-two two-point Pad\'e
approximant}. Its broadband error is close to the best possible for four elements with these exact
limits; the base network is the most accurate at low frequencies. A lower broadband error with four
elements requires giving up $Z(\infty)=1$, as in ladder (b) of Sec.~\ref{sec:term} (0.096).

\subsection{Broadband termination of the nested ladder}\label{sec:term}
The nested ladder spends all its elements on the low-frequency side. The exact \emph{tail}, i.e.\
the remainder of the infinite ladder behind section $n$, becomes purely resistive at high frequency.
This suggests keeping the nested core and adapting only the last section, a two-point idea in its
simplest form.

\para{(a) Last resistor from $Z(\infty)=1$.}At $s\to\infty$ compliances are short circuits and
masses are open, so the ladder reduces to the resistive ladder
$Z(\infty)=1/(G_0+1/(R_1+1/(G_2+\dots)))$. The value of the last resistive element that makes
$Z(\infty)=1$ follows from a short real recursion. For $N=4$ this gives $R_1=128/(128-9\pi^2)$: the
base network of Sec.~\ref{sec:family}.

\para{(b) Last two elements fitted.}The reactive and the resistive element of the last section are
fitted (positive, logarithmic coordinates) to minimise the maximum error over $0.1\le ka\le50$; all
other elements stay nested.

\begin{figure}[htbp]
\centering
\includegraphics[width=0.95\textwidth]{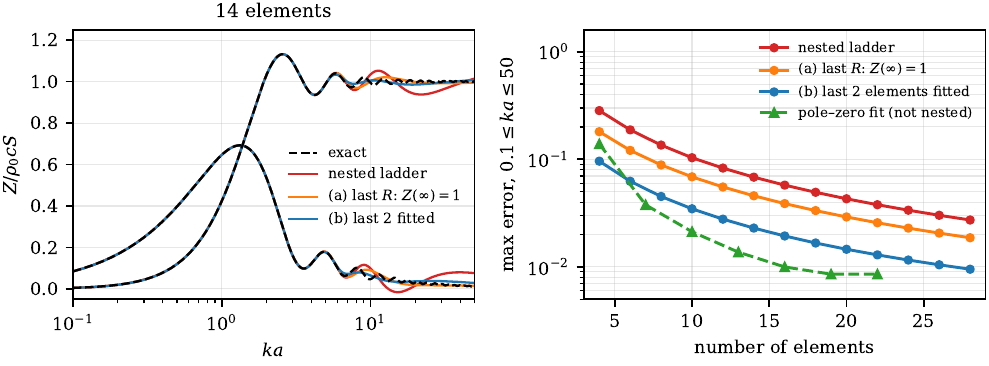}
\caption{Left: 14-element networks over $0.1\le ka\le50$. Right: broadband error versus number of
elements for the nested ladder, its terminated variants (a) and (b), and the non-nested pole--zero
fit.}\label{fig:term}
\end{figure}

Table~\ref{tab:term} and Figure~\ref{fig:term} show the effect. With 14 elements the broadband error
drops from 0.068 (nested) to 0.046 with (a) and to 0.023 with (b). Above $ka\approx10$ the remaining
error is mainly the decaying ripple of $Z$: for large $ka$, $2J_1(2ka)/2ka$ and the corresponding
Struve term oscillate with period $\Delta ka=\pi$ and amplitude $\approx\pi^{-1/2}(ka)^{-3/2}$,
about 0.02 at $ka=10$. A rational function of finite degree has only finitely many extrema on the
frequency axis and so cannot follow infinitely many oscillations; a network of given size follows
the ripple up to some $ka$ and averages over it beyond, which sets an error floor of the order of the
ripple amplitude there.

\begin{table}[htbp]
\caption{Maximum error over $0.1\le ka\le50$: nested ladder, variants (a) and (b), and the fitted
terminating elements of (b).}\label{tab:term}
\centering\small
\setlength{\tabcolsep}{4.5pt}
\begin{tabular}{rcccl}
\toprule
$N$ & nested & (a) & (b) & termination (b)\\
\midrule
4 & 0.284 & 0.181 & 0.096 & $C=0.1415$, $R=2.6972$\\
6 & 0.188 & 0.121 & 0.063 & $M=0.5380$, $G=0.4447$\\
8 & 0.136 & 0.089 & 0.045 & $C=0.0853$, $R=2.0006$\\
10 & 0.104 & 0.069 & 0.035 & $M=0.3596$, $G=0.3652$\\
12 & 0.083 & 0.055 & 0.028 & $C=0.0611$, $R=1.7333$\\
14 & 0.068 & 0.046 & 0.023 & $M=0.2706$, $G=0.3264$\\
16 & 0.058 & 0.039 & 0.019 & $C=0.0477$, $R=1.5797$\\
18 & 0.049 & 0.033 & 0.017 & $M=0.2163$, $G=0.3023$\\
20 & 0.043 & 0.029 & 0.015 & $C=0.0385$, $R=1.4907$\\
22 & 0.038 & 0.026 & 0.013 & $M=0.1772$, $G=0.2884$\\
24 & 0.034 & 0.023 & 0.012 & $C=0.0325$, $R=1.4242$\\
26 & 0.030 & 0.021 & 0.010 & $M=0.1519$, $G=0.2768$\\
28 & 0.027 & 0.019 & 0.010 & $C=0.0281$, $R=1.3738$\\
\bottomrule
\end{tabular}
\end{table}

\subsection{Other constructions}\label{sec:alt}
\para{Fitted Foster-type network.}A Foster-II admittance of a shunt mass $M_0$, a resistance
$R_0=128/9\pi^2$, a series branch $R_\infty$ with $C_1$, and $m$ parallel series R--M--C branches is
passive for any positive values ($N=4+3m$ elements). For $m=0$ and $R_\infty=128/(128-9\pi^2)$ it is
the four-element family of Sec.~\ref{sec:family}; fitting $C_1$ gives its optimum (\pzFour). For
$m\ge1$ the poles of the remaining admittance are found by vector fitting~\cite{gustavsen,sanathanan};
each pole pair $-\sigma\pm\mathrm{j}\omega$ becomes a branch with $R=2\sigma M$ and $C=1/M|p|^2$, and
all values are refined in logarithmic coordinates towards the minimax error. Over
$0.1\le ka\le50$ this gives \pzSeven, \pzTen, \pzThirteen\ and \pzSixteen\ with 7, 10, 13 and 16
elements; fits optimised for $ka\le5$ were computed in the same way up to 22 elements ($m=6$). The optimisation is non-convex: each order is started both from vector fitting and from
the previous order with one branch added or the next with one removed, and the best result is kept.
The element values change with $m$; the network is not nested. Because the optimisation is
non-convex, the fitted errors quoted here are upper bounds: a longer search can still find lower
minima, but never higher ones.

\subsection{All methods compared}\label{sec:compare}
Table~\ref{tab:methods} compares the methods of this tutorial in two
error measures: the absolute error $|Z-Z_\mathrm{exact}|/\rc$ over $0.1\le ka\le50$ used so far, and
the relative error $|Z/Z_\mathrm{exact}-1|$ over $0.005\le ka\le50$, which also weighs the very low
frequencies where $|Z|\to0$.

\begin{table}[htbp]
\caption{Approximations of the piston impedance. Elements: number of R, M, C; for rational
approximants that were not synthesised, the degree. ``PR'': positive real (checked), realisable by
Brune or Bott--Duffin synthesis but not as a ladder of the same size. Errors of fitted networks
are upper bounds (non-convex optimisation); Thiele and AAA use log-spaced nodes and samples.}\label{tab:methods}
\centering\small
\begin{tabular}{lcccll}
\toprule
method & elements & abs.\ error $0.1..50$ & rel.\ error $0.005..50$ & nested & passive\\
\midrule
nested ladder & 4 & 0.284 & 0.290 & yes & yes\\
nested ladder & 14 & 0.068 & 0.068 & yes & yes\\
ladder (a) & 14 & 0.046 & 0.045 & all but 1 & yes\\
ladder (b) & 4 & 0.096 & 0.099 & all but 2 & yes\\
ladder (b) & 14 & 0.023 & 0.023 & all but 2 & yes\\
\midrule
base network = ladder (a) & 4 & 0.181 & 0.187 & all but 1 & yes\\
two-point Pad\'e, order 2 & 4 & 0.153 & 0.136 & -- & yes\\
Beranek & 4 & 0.156 & 0.138 & -- & yes\\
minimax optimum of the family & 4 & 0.139 & 0.135 & -- & yes\\
\midrule
two-point Pad\'e, order 1 ($R\parallel M$) & 2 & 0.320 & 0.271 & -- & yes\\
two-point Pad\'e, order 3 ($k=4$) & deg 3 & 0.099 & 0.094 & no & PR\\
two-point Pad\'e, order 4 ($k=7$) & deg 4 & 0.070 & 0.069 & no & PR\\
two-point Pad\'e, order 5 ($k=8$) & deg 5 & 0.049 & 0.047 & no & PR\\
\midrule
pole--zero fit (Foster II) & 4 & 0.139 & 0.135 & no & yes\\
pole--zero fit (Foster II) & 13 & 0.014 & 0.014 & no & yes\\
AAA, degree 7 & deg 7 & 0.011 & 0.011 & no & not guaranteed\\
Thiele, 9 log-spaced nodes & deg 4 & 0.059 & 0.060 & no & not guaranteed\\
\midrule
Struve approximation~\cite{aj2003} & 3 const. & 0.0016 & 0.0014 & -- & --\\
\bottomrule
\end{tabular}
\end{table}

The results illustrate each lesson of the earlier sections.
\begin{itemize}
\item \emph{Taylor series.} Although the series converges for every $s$, its truncations are
useless as broadband approximations (relative error $10^{12}$ when truncated after $s^9$,
Sec.~\ref{sec:res}).
\item \emph{One-point Pad\'e at 0}, i.e.\ the nested ladder, is excellent at low frequency and
improves without bound with the order, but at low order levels off at the wrong value at high
frequency.
\item \emph{Two-point Pad\'e} matches the mass and radiation resistance at low frequency and the
limit $\rc$ with its $1/s$ correction at high frequency. Every two-point approximant in the table is
positive real with all poles in the left half-plane. Orders 1 and 2 are directly buildable networks;
beyond order 2 its error improves only gradually, because the maximum sits in the mid band, where the
ripples live.
\item \emph{Fits} (vector fitting, AAA~\cite{aaa}) reach the smallest broadband errors at a given
size; Thiele interpolation is less reliable. All depend on the sample points, are not nested, and except for the Foster-type
construction do not guarantee passivity.
\item \emph{Factoring out the known structure} wins outright when no network is needed: the
approximation $\Ho(x)\approx\frac2\pi-J_0(x)+\big(\frac{16}\pi-5\big)\frac{\sin x}x+
\big(12-\frac{36}\pi\big)\frac{1-\cos x}{x^2}$ of~\cite{aj2003}, combined with the exact real part,
gives 0.14\% over the whole range with three fixed constants. This is remedy (2) of
Sec.~\ref{sec:beyond} in action.
\end{itemize}

\para{Choice of network.}This confirms the summary in Sec.~\ref{sec:intro}. Over $0.1\le ka\le50$ the fitted
Foster-type network reaches \pzThirteen\ with 13 elements (Figure~\ref{fig:term}); up to $ka=5$ it
and the 14-element ladder are about equally accurate ($\pzThirteenFive$ and $\ladFourteenFive$), but
with 22 elements the ladder reaches $9.8\times10^{-9}$ against $\pzTwentytwoFive$ for the fit.

\section{Case study 2: the throat impedance of a horn}\label{sec:horn}
A horn loudspeaker presents its driver with the \emph{throat} impedance of the horn, which depends on
the horn contour and on the radiation load at the mouth. McLean, Post and Hixson~\cite{mclean} compute it
by cutting the horn into $N$ short uniform waveguide sections, multiplying their transmission matrices,
and terminating the product with the radiation impedance of a piston in a baffle at the mouth. This
example connects to the present paper in two ways: the cascade \emph{is} a continued fraction, and its
tail is the piston impedance for which Sec.~\ref{sec:piston} supplies networks. More detailed
models replace the plane waves by mode matching between duct sections and add edge diffraction at a
finite baffle~\cite{kolbrek}, or replace the plane mouth by a pulsating portion of a
sphere~\cite{helie}; the plane-wave cascade with a piston mouth is used here because it shows the
continued-fraction structure most directly.

\subsection{The cascade is a continued fraction}\label{sec:horncf}
A uniform section of length $l$, area $S_j$ and characteristic impedance $Z_j=\rho_0c/S_j$ relates
pressure and volume velocity at its two ends by
\begin{equation}
\begin{pmatrix}p_{j-1}\\U_{j-1}\end{pmatrix}
=\begin{pmatrix}\cos kl & \mathrm{j}Z_j\sin kl\\ \mathrm{j}\sin kl/Z_j & \cos kl\end{pmatrix}
\begin{pmatrix}p_j\\U_j\end{pmatrix},
\label{eq:abcd}
\end{equation}
and the throat impedance follows from the product of all matrices as
$Z_\mathrm{t}=(AZ_\mathrm{m}+B)/(CZ_\mathrm{m}+D)$, with $Z_\mathrm{m}$ the mouth
impedance~\cite[Eqs.~(8)--(9)]{mclean}. Each matrix is a M\"obius map
$Z_{j-1}=(A_jZ_j+B_j)/(C_jZ_j+D_j)$, and a product of such maps, evaluated from the mouth to the
throat, is exactly the backward recurrence of a continued fraction (Sec.~\ref{sec:nutshell}). It
becomes the familiar form when each section is replaced by its lumped equivalent, a series inertance
$M_j=\rho_0l/S_j$ and a shunt compliance $C_j=S_jl/\rho_0c^2$. With T-sections (half the mass on either
side of each compliance, adjacent halves merged),
\begin{equation}
Z_\mathrm{t}=sM_1+\cfrac{1}{sC_1+\cfrac{1}{sM_2+\cdots+\cfrac{1}{sC_N+\cfrac{1}{sM_{N+1}+Z_\mathrm{m}}}}}\,,
\label{eq:hornladder}
\end{equation}
with $s=\mathrm{j}\omega$: the horn is a ladder in physical space, a low-pass chain of masses and
compliances. Its tail $Z_\mathrm{m}$ is the radiation impedance, so inserting the nested ladder of
Sec.~\ref{sec:lad} for $Z_\mathrm{m}$ turns horn plus radiation into one passive ladder of positive
constant elements, ready for a circuit simulator. In the limit $l\to0$ the cascade solves Webster's
horn equation~\cite{webster,mclean}.

\subsection{The infinite exponential horn}\label{sec:hornexp}
For an exponential horn, $S(x)=S_\mathrm{t}\mathrm{e}^{mx}$, Webster's equation
$p''+mp'+k^2p=0$ has closed-form solutions, with cut-off $k_\mathrm{c}=m/2$, i.e.\
$f_\mathrm{c}=mc/4\pi$. For the outgoing wave of an infinite horn, the normalised throat impedance is,
with $\sigma=2\mathrm{j}k/m=\mathrm{j}f/f_\mathrm{c}$,
\begin{equation}
\frac{Z_\mathrm{t}}{\rho_0c/S_\mathrm{t}}=z(\sigma)=\frac{\sigma}{1+\sqrt{1+\sigma^2}},
\label{eq:zexp}
\end{equation}
whose real part above cut-off is the classical $\sqrt{1-(f_\mathrm{c}/f)^2}$. Because
$\sigma^2=(\sqrt{1+\sigma^2}-1)(\sqrt{1+\sigma^2}+1)$, the function satisfies $z=1/(2/\sigma+z)$, so
\begin{equation}
z(\sigma)=\cfrac{1}{\dfrac{2}{\sigma}+\cfrac{1}{\dfrac{2}{\sigma}+\cfrac{1}{\dfrac{2}{\sigma}+\cdots}}}\,,
\label{eq:zexpcf}
\end{equation}
a \emph{periodic} continued fraction. Read as a ladder, it alternates shunt inductances and series
capacitances, all equal to $\tfrac12$ in normalised units: the exponential horn is exactly an infinite
high-pass filter with its cut-off at $f_\mathrm{c}$. Unlike the piston fraction, this one does not converge everywhere on
the frequency axis. Below cut-off ($\sigma$ on the imaginary axis, $|\sigma|<1$) the fraction converges
geometrically (error below $\cfInfStop$ after 60 levels). In the passband, where the two fixed points
of $z\mapsto1/(2/\sigma+z)$ have equal modulus~\cite{wall}, the truncations do not converge at all: at
$f=2.5f_\mathrm{c}$ the error of the 40th to 80th truncation wanders between $\cfInfPass$ and
$\cfInfPassMax$. Slightly off the axis ($\sigma=0.2+2.5\mathrm{j}$) the error is still $\cfInfOff$
after 40 and $\cfInfOffEighty$ after 80 levels. On the frequency axis a truncated ladder therefore
needs a termination, here the exact tail $z$ itself (the image impedance), just as the piston ladder
needs the terminations of Sec.~\ref{sec:term}.

\subsection{A finite horn with a piston mouth}\label{sec:hornfinite}
The JBL 2360 horn of~\cite{mclean} is specified only graphically, so we use an exponential horn of the
same size: throat radius $\hornRt$\,mm (a 2-inch driver exit), length $\hornL$\,m, mouth radius
$\hornRm$\,cm. Then $m=\hornM$\,m$^{-1}$, $f_\mathrm{c}=\hornFc$\,Hz, the area ratio is $\hornAreaRatio$,
and at cut-off the mouth has $ka=\hornKaFc$. The mouth load is the circular piston~\eqref{eq:Z} of
radius $a=\hornRm$\,cm, normalised to the throat as
$Z_\mathrm{m}=(\rho_0c/S_\mathrm{m})\,Z(ka)/\rc$; McLean et al.\ use a rectangular piston, which behaves
similarly for aspect ratios near one~\cite{mclean}. The band is 50--5000\,Hz, so the mouth sees
$\hornKaMin\le ka\le\hornKaMax$. The reference is the exact transmission matrix of the exponential
section, built from the closed-form solutions of Webster's equation; terminated by the infinite-horn
impedance it reproduces~\eqref{eq:zexp} at the throat to within $\hornMatchedCheck$. All errors below
are $\max|\Delta Z_\mathrm{t}|$ over the band, normalised to $\rho_0c/S_\mathrm{t}$.

\para{Model error and approximation error.}Two kinds of error must be kept apart. The
\emph{acoustic model}, plane waves in a rigid exponential horn without losses (unless stated), a
circular piston in an infinite baffle at the mouth, determines how well $Z_\mathrm{t}$ describes a
real horn; it is not assessed here, and measurements~\cite{mclean,kolbrek} or more detailed
models~\cite{kolbrek,helie} are needed for that. The \emph{approximation error}, studied below,
is the deviation from the exact solution of this same model, caused by discretising the horn and by
replacing the piston impedance with a network. Only the latter is a property of the
continued-fraction and network method; the numbers below say nothing about the former.

\begin{figure}[htbp]
\centering
\includegraphics[width=0.95\textwidth]{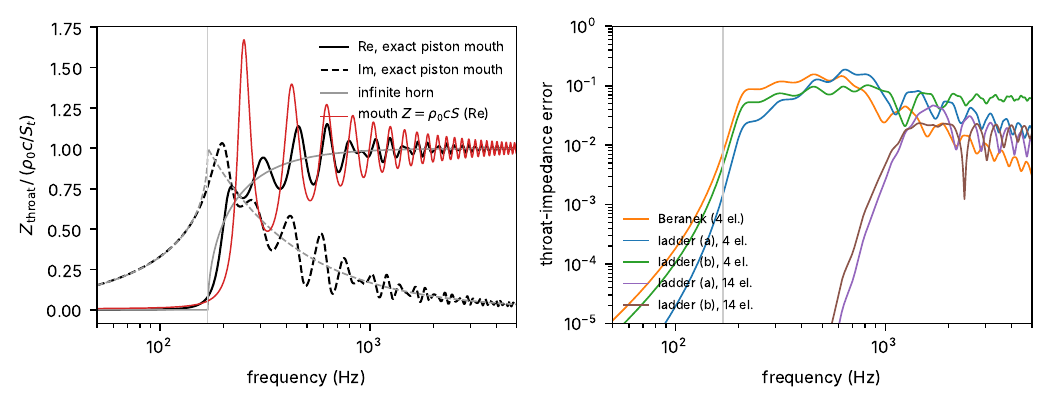}
\caption{Exponential horn ($f_\mathrm{c}=\hornFc$\,Hz, vertical line) with a piston mouth. Left: throat
impedance with the exact piston load, compared with the infinite horn~\eqref{eq:zexp} and with a
plane-wave mouth load $\rho_0c/S_\mathrm{m}$. Right: error of the throat impedance when the mouth is
modelled by networks of Sec.~\ref{sec:piston}.}
\label{fig:horn}
\end{figure}

\para{Throat impedance.} Figure~\ref{fig:horn} (left) shows the well-known behaviour: below cut-off the
throat is a mass, and above it the impedance oscillates around the infinite-horn curve because the
mouth reflects part of the wave. The ripple depends strongly on the mouth model: terminating the mouth
by its high-frequency limit $\rho_0c/S_\mathrm{m}$ (i.e.\ $Z=\rc$, the plane-wave row of
Table~\ref{tab:horn}) instead of the piston changes $Z_\mathrm{t}$ by up to $\htrhoc$. The
radiation load therefore has to be modelled; McLean et al.\ use the radiation impedance of a
rectangular piston in a baffle for it~\cite{mclean}.

\begin{figure}[htbp]
\centering
\includegraphics[width=0.62\textwidth]{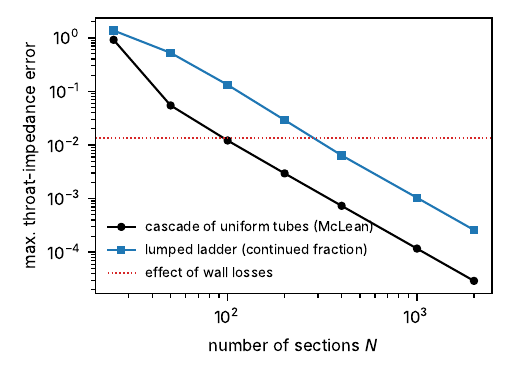}
\caption{Maximum throat-impedance error over 50--5000\,Hz versus the number of sections, for the cascade
of uniform tubes~\eqref{eq:abcd} and the lumped ladder~\eqref{eq:hornladder}. Dotted: size of the
thermoviscous losses.}
\label{fig:hornconv}
\end{figure}

\para{How many sections?} McLean et al.\ use 1000 sections of about 0.8\,mm. Figure~\ref{fig:hornconv}
shows that the staircase of uniform tubes converges with order $\stOrder$ in $N$: $\stHundred$ with 100
sections, $\stThousand$ with 1000, and $\stNeeded$ sections already give $10^{-3}$ over the whole band,
which spans $\hornLambdaRatio$ wavelengths at 5\,kHz. The lumped ladder~\eqref{eq:hornladder} also
converges with order $\ladOrder$, but with a ten times larger constant ($\ladHundred$ with 100,
$\ladThousand$ with 1000 sections); it needs $\ladNeeded$ T-sections for $10^{-3}$. The thermoviscous
wall losses of~\cite{mclean} (Kirchhoff boundary layer, via a complex wavenumber~\cite{pierce}) change
$Z_\mathrm{t}$ by at most $\lossMax$, near cut-off ($\lossFreq$\,Hz), and by less than $\lossHigh$ above
1\,kHz; the comparisons below are lossless.

\begin{table}[htbp]
\centering\small
\caption{Mouth modelled by a network: maximum error of the piston impedance itself (over the mouth's
$ka$ range, units $\rc$) and of the resulting throat impedance (units $\rho_0c/S_\mathrm{t}$), with the
frequency of the latter maximum.}
\label{tab:horn}
\setlength{\tabcolsep}{3.5pt}
\begin{tabular}{lccc}
\toprule
mouth model & piston err. & throat err. & at (Hz)\\
\midrule
plane wave ($Z=\rho_0cS$) & 0.989 & 0.986 & 248\\
Beranek (4 el.) & 0.156 & 0.155 & 457\\
ladder (a), 4 el. & 0.181 & 0.187 & 643\\
ladder (b), 4 el. & 0.096 & 0.102 & 824\\
nested ladder, 14 el. & 0.068 & 0.072 & 3972\\
ladder (a), 14 el. & 0.046 & 0.046 & 1692\\
ladder (b), 14 el. & 0.023 & 0.023 & 1457\\
\bottomrule
\end{tabular}
\end{table}

\para{Which mouth network?} Table~\ref{tab:horn} and Figure~\ref{fig:horn} (right) replace the exact
piston by networks of Sec.~\ref{sec:piston}. The result is simple: the error of the throat impedance
equals the error of the mouth network, to within a factor $\hornSens$ above $2f_\mathrm{c}$. Below
cut-off the error is small, because every network is accurate at low $ka$. The horn neither hides nor magnifies the radiation model, so the accuracies of
Secs.~\ref{sec:res}--\ref{sec:compare} carry over directly. Beranek's network gives $\htber$, the
four-element base network (ladder (a)) $\htlaFour$, and ladder (b) $\htlbFour$ with four and $\htlbFourteen$
with 14 elements. With ladder (b) and $\ladNeeded$ T-sections, horn and radiation together form one
passive ladder of $\totalElements$ positive elements~\eqref{eq:hornladder}. With transmission-line
elements instead of lumped sections, $\stNeeded$ lines suffice.

\para{Outlook.} The ladder~\eqref{eq:hornladder} is large but is a continued fraction with known
coefficients. Reducing it with the two-point Pad\'e reduction of Sec.~\ref{sec:rcline} would give a
small network for the throat impedance, exact at low frequency and at $Z_\mathrm{t}(\infty)$; whether
its elements stay positive, as they do for the RC line, is left for future work. Measured throat
impedances, as in~\cite{mclean,kolbrek}, are the test of the plane-wave and piston assumptions
themselves; for constant-directivity horns McLean et al.\ found agreement with the cascade model and
traced the resonances of the throat impedance to reflections at the diffraction
slot~\cite{mclean}.

\section{Discussion}\label{sec:disc}
The ladder of Sec.~\ref{sec:lad} combines the three desired properties with no optimisation at all,
with passivity certified up to 202 elements.
The only input is the Taylor series~\eqref{eq:taylor}; the construction is a single pass of the
Viskovatov algorithm followed by an even contraction. In the language of Secs.~\ref{sec:twopoint}
and~\ref{sec:networks} it is the network form of the diagonal one-point Pad\'e approximants, a mixed
Cauer ladder with one element per matched Taylor coefficient; it thus avoids the element overhead of
transformerless (Bott--Duffin) synthesis~\cite{bottduffin,hughes2017}.

\para{Status of the results.}Exact (closed form): the continued-fraction and ladder structure, the
elements $M_0$, $G_0$ and $C_1$~\eqref{eq:first}, and the four-element family including the two-point
approximant~\eqref{eq:order2}. Rigorously certified (interval arithmetic): $d_k,e_k>0$ for $k\le100$,
hence passivity up to 202 elements. Numerically verified (high-precision arithmetic): all errors,
bandwidths and the horn results; the fitted networks give upper bounds of what is achievable.
Conjectured: positivity for all $k$.

Three points deserve attention.
\begin{itemize}
\item \emph{Positivity} is proven for $k\le100$, which covers every practical order. A proof for all
$k$ requires explicit remainder bounds on~\eqref{eq:kasym}, or a moment representation of $Y=1/Z$ via
the delay integral~\eqref{eq:delay}.
\item \emph{Pad\'e character.} The expansion point is $s=0$, so $Z(\infty)$ is not enforced. The
terminations of Sec.~\ref{sec:term} repair this at the cost of one or two non-nested elements. A
fully nested two-point continued fraction (an M-fraction) that alternates between~\eqref{eq:taylor}
and~\eqref{eq:asym}, \emph{and} whose coefficients stay positive, is the natural next step; the
order-1 and order-2 two-point approximants of Sec.~\ref{sec:family} are its first members.
\item \emph{Numerics.} The Viskovatov recursion loses digits quickly. Extended precision is required
for high orders, but only once: the element values are then simply tabulated.
\end{itemize}
The same procedure applies to any radiation quantity with a known power series, such as the
higher-order Struve-function quantities of~\cite{aj2016} or the mutual impedance of coplanar pistons,
provided the positivity of the contracted coefficients is checked.

\section{Conclusion}
Continued fractions, Pad\'e approximants and ladder networks are three views of the same object.
For the baffled-piston radiation impedance the one-point C-fraction alternates in sign, but its even
contraction has positive coefficients (certified for the first 100 pairs, conjectured for all) and is
realised exactly as a ladder of series compliance--resistance and shunt mass--resistance sections.
The construction extends to any order without changing existing values; positivity, and hence
passivity, is certified up to 202 elements and conjectured beyond. The networks reproduce
the first $N$ Taylor coefficients with $N$ elements, and are valid up to $ka\approx0.58N-1.2$ at 1\%
accuracy. At four elements a one-parameter family unifies the classical networks; Beranek's fitted
network is, to about 2\%, the order-two two-point Pad\'e approximant, now available in closed form.
Adapting only the last one or two elements of the nested ladder gives good broadband networks,
and a fitted Foster-type network is somewhat more accurate over a wide band at low order but is not
nested. The same view covers horns: a cascade of waveguide sections is a continued fraction whose
tail is the radiation impedance, the infinite exponential horn is a periodic one (an exact
high-pass ladder), and a mouth network transfers its accuracy one to one to the throat impedance.

\section*{Acknowledgements}
The author is grateful to
\href{https://www.linkedin.com/in/ren\%C3\%A9-christensen-87b5a15/}{Ren\'e Christensen}
(Acculution ApS), whose LinkedIn post of 1~October 2026 rekindled the author's interest in the
radiation impedance of a circular piston in an infinite, planar, rigid baffle, a topic on which the
author first published in 2003~\cite{aj2003}.

\section*{Funding}
The author received no funding to complete this research.

\section*{Conflict of interest}
The author declares no conflict of interest.

\section*{Author contributions}
R.\,M. Aarts: conceptualization, methodology, software, validation, formal analysis, writing.

\section*{Use of AI-assisted technology}
The AI assistant Claude (Anthropic; model Claude Opus 5.5) was used to assist with derivations, with language corrections and
checking the Python code, with numerical verification. All content was reviewed and verified by the
author, who takes full responsibility for it.

\end{document}